\documentclass{elsart1p}
\usepackage{graphicx}
\usepackage{amssymb}
\begin{document}
\begin{frontmatter}
%
%
%
\title{PHENIX results on the $\sqrt{s_\mathrm{NN}}$ dependence of jet~quenching}
%
%
\author{Christoph Baumann, for the PHENIX Collaboration}
\address{University of Muenster, Wilhelm-Klemm-Str. 9, D-48149 Muenster, Germany}
\begin{abstract}

The PHENIX experiment has established jet quenching at a center-of-mass energy of 200~GeV in Au+Au collisions. Recent measurements of Cu+Cu collisions at the same center-of-mass energy support a parton energy loss scenario. Furthermore, the onset of jet quenching was studied in Cu+Cu collisions for three center-of-mass energies 22.4, 62.4 and 200~GeV.

\end{abstract}
\begin{keyword}
%
\PACS
\end{keyword}
\end{frontmatter}
%
\section{Introduction}
First results of the PHENIX experiment have established the suppression of high-$p_\mathrm{T}$ hadrons in Au+Au collisions at $\sqrt{s_\mathrm{NN}} = 200\mathrm{~GeV}$~\cite{Adcox:2004mh}. The suppression can be quantified relative to a p+p reference with the nuclear modification factor $R_\mathrm{AA} = \frac{(d^2N/(dp_\mathrm{T}dy))_\mathrm{AA}}{T_\mathrm{AA}\cdot(d^2\sigma/(dp_\mathrm{T}dy))_\mathrm{pp}}$.
The $R_\mathrm{AA}$ of direct photons does not show a suppression for $p_\mathrm{T} \leq 10-12\mathrm{~GeV/c}$, neither do hadrons in d+Au collisions.
Together, this is commonly interpreted as evidence for energy loss of partons in a hot and dense medium which is produced in ultra-relativistic heavy ion collisions and is observed as jet quenching. With the data sets up to Run 6 obtained at PHENIX, it has been possible to extend the $p_\mathrm{T}$ coverage of the data and to examine the dependence on the colliding particle species as well as on the center-of-mass energy.

\section{Results}
\begin{figure}
\begin{center}
\includegraphics[width=\textwidth]{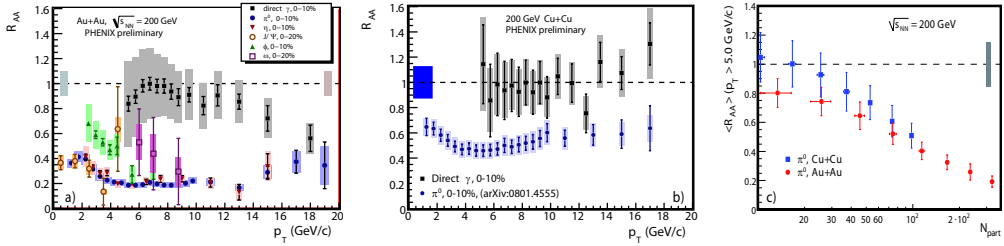}
\caption{a) Nuclear modification factors for different mesons in central Au+Au collisons at $\sqrt{s_\mathrm{NN}} = 200\mathrm{~GeV}$. b) Nuclear modification factor for direct photons and $\pi^0$ in central Cu+Cu collisions. c) Comparison of the nuclear modification factor for $\pi^0$ in Au+Au and Cu+Cu collisions as a function of $N_\mathrm{part}$. }
\label{img:raa-all}
\end{center}
\end{figure}

\begin{figure}
\begin{center}
\includegraphics[width=\textwidth]{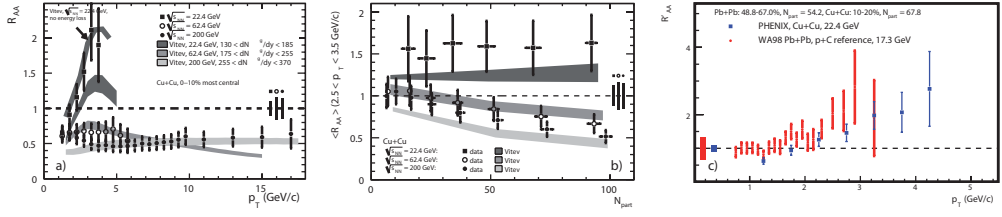}
\caption{a) nuclear modification factor for the most central events in Cu+Cu collisions at 22.4, 62.4 and 200~GeV, grey bands show theoretical calculations. b) Comparison of the integrated $R_\mathrm{AA}$ as a function of the number of participants. c) Comparison of the PHENIX results at 22.4~GeV in Cu+Cu collisions to the SPS results at 17.3~GeV in Pb+Pb collisions relative to a p+C reference for similar $N_\mathrm{part}$ values}
\label{img:CuCuEDep}
\end{center}
\end{figure}

The new PHENIX results extend the neutral pion spectra in Au+Au collisions at $200\mathrm{~GeV}$ to $p_\mathrm{T}$ up to 20~GeV/c, for direct photons up 18~GeV/c. Moreover, the $R_\mathrm{AA}$ has been determined for a large variety of mesons.
For neutral pions, the nuclear modification factor is approximately constant at $\approx 0.2$ for $p_\mathrm{T} \geq 5\mathrm{~GeV/c}$~\cite{Adare:2008qa}, $\eta$ mesons show the same behavior, as shown in Figure~\ref{img:raa-all}a. Within errors also the $R_\mathrm{AA}$ of the $J/\Psi$, measured at midrapidity, is compatible with the $\pi^0$ and $\eta$ $R_\mathrm{AA}$.
However, the $R_\mathrm{AA}$ of the $\omega$ and the $\phi$ meson is only $\approx 0.5$ at $p_\mathrm{T}= 6-8 \mathrm{~GeV/c}$ . This deviating behavior will be important to understand in detail and is a promising tool for testing theoretical models.
The nuclear modification factor for the direct photons is below unity for $p_\mathrm{T} \geq 13\mathrm{~GeV/c}$ in Au+Au collisions. This was not observed in the Cu+Cu data (Fig.~\ref{img:raa-all}b). The photon suppression in Au+Au can be explained by initial state effects, such as an isospin effect or the EMC effect, or by the suppression of fragmentation photons. The suppression of neutral pions in the most central Cu+Cu collisions amounts to a factor of 2. A comparison to the Au+Au data (Fig~\ref{img:raa-all}c) shows, that the suppression follows a simple $N_\mathrm{part}$ dependence, which is consistent with the assumption of a $N^{2/3}_\mathrm{part}$ dependence of partonic energy loss~\cite{Adare:2008qa,Vitev:2005he}. The available Cu+Cu data allow to chart the center-of-mass energy dependence of jet quenching from the SPS energy regime up to top RHIC energies. Three data-sets are available, measured at 22.4~GeV, 62.4~GeV and 200~GeV.
For the latter two, a p+p reference measured by PHENIX is also available, for the 22.4~GeV data, we use a parameterization of world data as a reference~\cite{Arleo:2008zd}. The nuclear modification factors for the most central class are compared in Figure~\ref{img:CuCuEDep}a~\cite{:2008cx}.
For $p_\mathrm{T} > 5\mathrm{~GeV/c}$, the suppression of the 62.4~GeV data is compatible with the 200~GeV data. The smaller energy loss in the 62.4~GeV data is presumably compensated for by steeper parton $p_\mathrm{T}$ spectra.
The 22.4~GeV data show no suppression, the nuclear modification factor is above unity. This can be attributed to the Cronin enhancement. It is conceivable that the Cronin enhancement hides a possible suppression. Figure~\ref{img:CuCuEDep}a shows theoretical scenarios, one with Cronin enhancement only and one with Cronin enhancement and jet quenching. The data do not allow to discriminate between these two. In Figure~\ref{img:CuCuEDep}b, the dependence of $R_\mathrm{AA}$ on $N_\mathrm{part}$ is shown for the three energies. While the suppression increases for more central events at 62.4~GeV and 200~GeV, the 22.4~GeV data do not exhibit a strong dependence on $N_\mathrm{part}$. This can either be explained by a weak centrality dependence of the Cronin enhancement or by a partial cancellation of the Cronin enhancement and jet quenching.
WA98 has recently reported on a suppression signal in Pb+Pb collisions at 17.3~GeV~\cite{Aggarwal:2007gw}. The suppression was observed in the most central collisions relative to a p+Pb and a p+C reference measured at 17.4~GeV and normalized to the respective $N_\mathrm{coll}$ values. For more peripheral centrality classes, the nuclear modification factor is above unity, similar to what is seen in the PHENIX data at 22.4~GeV. In Figure~\ref{img:CuCuEDep}c), we compare the WA98 data to the PHENIX results measured at 22.4~GeV, choosing two centrality classes with a similar $N_\mathrm{part}$. Within errors, the nuclear modification factors are compatible.
\section{Conclusions}
The flattening of $\pi^0$ $R_\mathrm{AA}$ at high $p_\mathrm{T}$, the suppression pattern for different mesons, the $T_\mathrm{AA}$ scaling of direct photons, the $N_\mathrm{part}$ dependence of the suppression as well as the dependence on the center-of-mass energy measured by PHENIX support an energy-loss picture. However, the different behavior of the direct photon $R_\mathrm{AA}$ at the highest $p_\mathrm{T}$ as well as the difference in the suppression of the $\omega$ and $\phi$ mesons compared to the $\pi^0$ and $\eta$ have to be understood in detail.
The consistency between the 22.4~GeV data and the WA98 results on neutral pion production suggests that the $N_\mathrm{part}$ dependence of the nuclear modification factor observed at 200 GeV is also valid in the SPS energy regime.
The available data show that jet quenching starts to prevail over the Cronin enhancement between 22.4~GeV and 62.4~GeV.

\label{}
%
%
%

%
\end{document}